\documentclass[pdflatex]{sn-jnl}% Basic Springer Nature Reference Style/Chemistry Reference Style
\usepackage{graphicx}
\usepackage{multirow}
\usepackage{subfigure}
\usepackage{microtype}

\usepackage[authoryear]{natbib}
\jyear{2023}%

\theoremstyle{thmstyleone}%
\theoremstyle{thmstyletwo}%

\theoremstyle{thmstylethree}%

\begin{document}

\title[How Social Relationships Affect Peer Assessment in E-Learning]{Impact of Social Relationships on Peer Assessment in E-Learning}

%%=============================================================%%
%% Prefix	-> \pfx{Dr}
%% GivenName	-> \fnm{Joergen W.}
%% Particle	-> \spfx{van der} -> surname prefix
%% FamilyName	-> \sur{Ploeg}
%% Suffix	-> \sfx{IV}
%% NatureName	-> \tanm{Poet Laureate} -> Title after name
%% Degrees	-> \dgr{MSc, PhD}
%% \author*[1,2]{\pfx{Dr} \fnm{Joergen W.} \spfx{van der} \sur{Ploeg} \sfx{IV} \tanm{Poet Laureate} 
%%                 \dgr{MSc, PhD}}\email{iauthor@gmail.com}
%%=============================================================%%

%\author{\fnm{Anonymous authors}}

\author*[1]{\fnm{Francisco} \sur{Sousa}}\email{francisco.sousa.98@tecnico.ulisboa.pt}

\author[1,2]{\fnm{Tomás} \sur{Alves}}\email{tomas.alves@tecnico.ulisboa.pt}

\author[1,2]{\fnm{Sandra} \sur{Gama}}\email{sandra.gama@tecnico.ulisboa.pt}

\author[1,2]{\fnm{Joaquim} \sur{Jorge}}\email{jorgej@tecnico.ulisboa.pt}

\author[1,2]{\fnm{Daniel} \sur{Gonçalves}}\email{daniel.j.goncalves@tecnico.ulisboa.pt}

\affil*[1]{\orgdiv{Instituto Superior Técnico}, \orgname{University of Lisbon}, \orgaddress{\street{Av. Rovisco Pais 1}, \city{Lisbon}, \postcode{1049-001}, \country{Portugal}}}

\affil[2]{\orgdiv{INESC-ID}, \orgaddress{\street{R. Alves Redol 9}, \city{Lisbon}, \postcode{1000-029}, \country{Portugal}}}

%%==================================%%
%% sample for unstructured abstract %%
%%==================================%%

%\todo[inline]{RV2(1): I think it is important to also mention that you are examining the use of certain aids to counter the possible interpersonal effects of peer-assessment.}

\abstract{Peer assessment has been widely studied as a replacement for traditional evaluation, not only by reducing the professors' workload but mainly by benefiting students' \textcolor{black}{engagement} and learning.
Although several works successfully validate its accuracy and fairness, more research must be done on how students' pre-existing social relationships affect the grades they give their peers in an e-learning course.
We developed a Moodle plugin to provide the platform with peer assessment capabilities in forums and used it on an MSc course. \textcolor{black}{The plugin curated the reviewer set for a post based on the author's relationships and included rubrics to counter the possible interpersonal effects of peer assessment.} Results confirm that peer assessment is reliable and accurate for works with at least three peer assessments, although students' grades are slightly higher. The impact of social relationships is noticeable when students who do not like another peer grade their work consistently lower than students who have a positive connection. However, this has little influence on the final aggregate peer grade.
Our findings show that peer assessment can replace traditional evaluation in an e-learning environment where students are familiar with each other.}

\keywords{Peer assessment, E-learning, Fairness, Reliability, Relationships}

%%\pacs[JEL Classification]{D8, H51}

%%\pacs[MSC Classification]{35A01, 65L10, 65L12, 65L20, 65L70}

\maketitle

\section*{Funding}
This work was supported by national funds through Fundação para a Ciência e a Tecnologia (FCT) with references SFRH/BD/144798/2019 and under project UIDB/50021/2020 (DOI:10.54499/UIDB/50021/2020).

%\section*{Declaration}
%This paper or a similar version is not currently under review by a journal or conference. This paper is void of plagiarism or self-plagiarism as defined by the Committee on Publication Ethics and Springer Guidelines.

%\section*{Conflict of interest/Competing interests}
%The authors declare that they have no conflict of interest.

%\section*{Ethics approval}
%Not applicable.

%\section*{Consent to participate}
%Not applicable.

%\section*{Consent for publication}
%Not applicable.

%\section*{Availability of data and materials}
%The datasets generated during and/or analyzed during the current study are available from the corresponding author on reasonable request.

%\section*{Code availability}
%Not applicable.

%\section*{Authors' contributions}

%Omitted for the review process.

% Francisco Sousa: Conceptualization, Investigation, Methodology, Writing – original draft, and Writing – review \& editing.
% Tomás Alves, Sandra Gama, Daniel Gonçalves, and Joaquim Jorge: Conceptualization, Methodology, Supervision, and Writing – review \& editing.

\section{Introduction}\label{intro}
%Peer assessment is valid and cool and has a and b advantage;
Peer assessment has gained followers as an alternative model to traditional student evaluation. \textcolor{black}{Peer assessment consists of students with similar backgrounds judging each other's work \citep{topping_peer_1998,na2019quantitative}, promoting reflection and discovery of new understandings by finding the difference between others and themselves~\citep{chang2020integration}.
Recent meta-analyses showcase the significance of such mechanisms in teaching and learning nowadays~\citep{zheng2020role,li2020does,yan2022effects} by improving students' academic performance~\citep{black2009developing,yan2022effects} through pedagogical activities that facilitate learning~\citep{adachi2018academics,double2020impact}.}
Previous work has shown that peer assessment can obtain the same accuracy and fairness of grades one would get from the professor \citep{topping_peer_2009}. %Moreover, peer assessment takes some of the load off professors.
%Coupled with virtual learning environments, it allows for the easy integration of complements. These include gamification elements to attract students' attention in terms of academic development, increasing their motivation to continue interacting with the new content \citep{kuo_how_2016}, and positively impacting student's attitudes towards the lesson \citep{yildirim_effects_2017}.
%Previous work has shown that peer assessment can obtain the same accuracy and fairness of grades one would get from the professor \citep{topping_peer_2009}. Peer assessment takes some of the load off professors and, together with virtual learning environments, allows for the easy integration of complements like gamification elements to attract students' attention in terms of academic development, increasing their motivation to continue interacting with the new content \citep{kuo_how_2016}, having a positive impact on students' attitudes towards the lesson \citep{yildirim_effects_2017}.
Due to their scalability, e-learning courses popularised peer grading, especially in Massive Open Online Courses (MOOC). 
There, the traditional assessment approach, which consists of manual grading, would be unfeasible and costly for professors to complete within a reasonable period, particularly in settings with many students and assignments.
\textcolor{black}{However, recent literature reviews point towards a large research gap in the field. Few past studies focused on understanding how social factors may explain the variance observed in peer assessment environments~\citep{gamage2021peer,panadero2023systematic}.}

In most e-learning courses, students usually do not know each other. However, virtualizing traditional learning environments due to the COVID-19 pandemic~\citep{dhawan2020online} brought peer assessment to e-learning environments where students interact outside of classes and are familiar with each other from previous years.
Researchers must consider a peer assessment bias related to student relationships in this student familiarity scenario.
Furthermore, peer assessment is a social activity requiring a mutual trust relationship~\citep{panadero2016safe}.
To minimize the effect of social relationships between students, researchers argued that anonymizing uni- or bidirectionally the parties could improve the peer assessment~\citep{li2017role}.
A recent review shows mixed results of using anonymity to minimize interpersonal effects in peer assessment environments~\citep{panadero_empirical_2019}.
For instance, \citet{lin2018anonymous} shows that anonymity positively impacts providing more critical peer feedback and fosters different types of peer feedback compared to students in non-anonymous conditions.
Research also shows that anonymity provides more comfort and less peer pressure to the students~\citep{raes2015increasing,vanderhoven2015if,seifert2019online}.
In contrast, the meta-analysis by \citet{panadero_empirical_2019} and \citet{li2016peer} suggest that non-anonymity is better for increasing students' peer grading accuracy when compared to teachers' assessment.
These mixed results highlight the complexity of factors that are present in peer assessment.

The current state of the art led us to reconsider whether concealing the identity of peers is always beneficial to control interpersonal effects in peer assessment.
In particular, we believe that providing peer assessment aids such as criteria, rubrics, and training may help alleviate the interpersonal effects.
Other researchers have already considered social relationships through the nomination of intimate friends~\citep{azarnoosh_peer_2013} or prediction of relationships based on student interactions~\citep{he_peer_2015} but did not find solid and reliable results.
However, to the best of our knowledge, there is insufficient empirical evidence on whether using peer assessment aids helps minimize the interpersonal effects of peer assessment.
%Literature is poor on the subject. There is one work which measures relationships based on online interactions. We propose studying self-reported relationships with students training beforehand.
%Nevertheless, there is a poor empirical evidence on the subject.
%Nevertheless, there is insufficient empirical evidence on the subject.
%Both \cite{azarnoosh_peer_2013} and \cite{he_peer_2015} focused on the influence of social relationships on peer assessment. %While 
%\cite{azarnoosh_peer_2013} used a small traditional learning course where she asked students %in 
%at the beginning of the semester to nominate their three most intimate friends %, 
%\cite{he_peer_2015} adapted a blended learning course to collect the relationships between students based on the interactions in their virtual learning environment. The first found no significant difference but used a small class, and the results from the second were inconclusive.
In this work, \textcolor{black}{we aim to understand how the social relationships between students affect the quality and students' perceptions of the peer assessment environment in an e-learning course.
We collected multiple self-reported relationships through peer nominations, peer ratings throughout the semester, and students' perceptions of the peer assessment process.
Then, we analyzed whether it is possible to overcome the relationship bias in the peer assessment environment, considering that faculty provided rubrics to train the students' grading and where most students were previously familiar with each other.}
%Our results show that students' relationships bias the peer assessment regarding grade accuracy and fairness.
%Furthermore, we show that applying peer assessment training and a grader selection algorithm minimizes the friendship bias in the peer assessment process to the point where it can be ignored.

The remainder of this paper is organized as follows: First, we discuss the related work on peer assessment. Then, we present our research methodology, including the peer-grading environment, data collection, and processing. Next, we discuss how we analyzed the data, the results obtained, and the implications for designing future experiences. Finally, we present our conclusions and pointers for future work.

\section{Related Work}
%Peer Assessment definition, e-learning environments, accuracy, fairness, importance of training, number of graders, peer nominations and ratings, relationships influence.

According to \cite{topping_peer_1998}, peer assessment -- or peer grading -- consists of students with similar backgrounds judging each other's work, including the number, level, value, practicality, quality, success, and the result of their daily study, with substantial evidence that it can improve the effectiveness and quality of learning \citep{li2020does}.
\textcolor{black}{The most common method to determine the quality of the peer assessment is by measuring the differences between grades provided by students and faculty members~\citep{alfallay2004role,kulkarni2013peer,yan2022effects}.
For instance, we can compare} the grade given to exercises by professors with that given by peers to infer if peer assessment is \textit{accurate} -- high correlation between student- and professor-assigned scores -- in awarding the same score, and \textit{fair} -- consistency of scores given by multiple student graders -- to all the students involved.
Most literature agrees that peer assessment is sufficiently similar to a professor's evaluation and, therefore, accurate to use with students, concluding there is no significant difference in who is grading \citep{azarnoosh_peer_2013, luo_peer_2014, usher_peer_2018, he_peer_2015}.
By comparing the influence of different numbers of graders in peer assessment fairness, \cite{luo_peer_2014} found that more is best, recommending between three and five graders. \cite{cho_validity_2006} suggest four to six graders.
\textcolor{black}{Other researchers decided to tackle the subjectiveness bias by applying other strategies such as anonymity~\citep{bostock2000student,lin2018anonymous,hoang2022does}.}

\subsection{Anonymity in Peer Assessments}

Although primarily used in e-learning courses, like MOOC, peer assessment is progressively growing in traditional classes and blended learning courses, the integration of face-to-face with both online instruction \citep{graham_emerging_2013} and online learning experiences \citep{garrison_blended_2004}. These are composed of students who know each other, and due to peers being peers, inevitably form relationships, resulting in bias when awarding grades.

% Talk about anonymization

There are several approaches to account for relationship biases in peer assessment.
For instance, researchers argue that anonymizing through single- or double-anonymized methods improves peer assessment (e.g.,~\citet{li2017role}).
\textcolor{black}{Past research (e.g.,~\citet{raes2015increasing,vanderhoven2015if}) found that anonymity allows students to overcome inhibitions and improve their evaluation skills.
Initial work by \citet{howard2010anonymity} reported that anonymizing the peer assessment led students to be five times more likely to create critical feedback and four times more likely to provide justifications for the
improvements they suggested than those in a non-anonymous group.
Later, \citet{guler2017use} found that an anonymous group provided peer ratings more correlated with the instructor ratings than a non-anonymized group, and \citet{gamage2017improving} showed that non-anonymized reviewers resulted in improved feedback and interaction in the peer assessment process.
More recent work by \citet{van2019effects} showed that an anonymous group of reviewers processed more directive higher-order feedback (e.g., feedback on ideas, organization, and argumentation) and obtained higher scores on their revised essays than a non-anonymous group.}

\textcolor{black}{Regarding students' perceptions, \citet{lin2018anonymous} investigated the role of anonymity in an online peer assessment within a Facebook-based learning application.
The authors leveraged two groups of students. One group had the assessors' identities hidden throughout the peer-assessment process, and the other showed the full real names of the graders to their reviewers.
Results show that anonymity increased cognitive comments but reduced affective comments.
\citet{lin2018anonymous} also observed that the anonymous group had a more positive attitude toward the single-anonymized system, particularly by reporting a higher level of perceived learning.
However, the authors reported that the anonymous group had a lower perception that peer assessment was fair.
Further work by \citet{seifert2019online} compared self- and anonymous peer-assessment to understand if these strategies encourage students to take more responsibility for the learning process.
Results showed that anonymity also provides students with more comfort and less peer pressure.
Recent results by \citet{kumar2019anonymous} also show that students had a favorable perception and attitude towards the online peer assessment method when the assessor and assessee are kept anonymous.
More recently, \citet{su2023masked} found that students preferred an anonymous peer assessment since it provided more comfort than a non-anonymous approach.}

Nevertheless, recent meta-analyses show mixed results of using anonymity to minimize interpersonal effects in peer assessment environments.
In particular, \citet{panadero_empirical_2019} and \citet{li2016peer} suggest that non-anonymity is better for increasing students' peer grading accuracy when compared to teachers' assessment.
These mixed results highlight the complexity of factors present in peer assessment and, in particular, led researchers to opt for considering relationship biases in the peer assessment rather than trying to remove them.

\subsection{Relationship Bias in Peer Assessments}

\textcolor{black}{Social relationships can be looked at more closely to find a friendship bias in peer assessment.}
Peer ratings -- rating friends on a Likert scale -- and peer nominations -- classify a small number of peers in a group by who they like the most and the least -- are sociometry tools to study and infer student relationships \citep{coie_dimensions_1982}.
\cite{azarnoosh_peer_2013} asked each student to nominate their three most intimate friends in class and compared friends' awarded grades with acquaintances. It found no significant difference but justified the result with the small class of 26 students who all know each other.

\cite{he_peer_2015} went further with more significant class sizes and, instead of only associating student relationships with peer assessments, tried to reduce the possible bias. \cite{he_peer_2015} tried to take advantage of interactions in the online discussion board to connect students and reassign graders to distant acquaintances since the course was primarily remote. A control group was left untouched for comparison. Although students reported they did not grade friends' exercises, the result comparison with the ones who did is inconclusive. Inferring relationships from online data is valid but does not accurately reflect students' lives offline.

\textcolor{black}{State-of-the-art research points towards an effect of friendship as a social factor on peer assessment.
In particular, researchers reported this bias affecting peer assessment scores or students' perceptions of the peer assessment (e.g., \citet{harris2013opportunities,dominguez2016comparative,kilickaya2017peer,ersoz2018facebook}).}
The primary concern is over-scoring based on friendship biases~\citep{panadero2013impact} since students believe that they may lose friendships if they provide poor grades~\citep{kilickaya2017peer}.
\citet{panadero2013impact} considered using rubrics to counter this bias, but this approach only reduced biases of low and moderate-level friendships.
In particular, high relationship levels produced significantly more over-scoring than low relationship levels between students.
Nevertheless, the state-of-the-art provides limited knowledge regarding the effect of relationship biases in peer assessment and, more precisely, how to assess these relationships between students.
Therefore, leveraging student relationships with peer nominations and peer ratings from a peer assessment environment may provide more robust and broader insights to help us understand the effect of the assessment relationship bias in an e-learning course.

\section{Research Methodology}
\textcolor{black}{This section describes the research questions tackled in the experiment, the remote learning course, the peer assessment environment, and the data collection and analysis.}

\subsection{Research Questions}

\textcolor{black}{Our objective is to understand whether student relationships affect the peer assessment environment in an e-learning course.
However, we need to tackle the quality of the peer assessment considering the tailoring of assessors and whether students perceived a bias based on the relationships.
We decided to run an experiment in an e-learning course for a semester.
At the beginning of the course, students anonymously self-reported their relationships with each peer by stating which colleagues were their most and least favorite.
Along the course, we also tracked how relationships between peers who had no previous relationship before the course emerged and considered them in the peer assessment environment dynamics.
During the course, students peer-assessed their colleagues' posts in a single-anonymized approach, i.e., only the reviewer knew the identity of the post's author.
To understand the role of social relationships in this process, we specifically choose the reviewers of a post based on the relationship between the post's author and their peers.
The faculty provided rubrics to train each student in the peer assessment to assist assessors in judging the quality of student performance.}

\textcolor{black}{In this light, a significant component of the peer assessment environment was tailoring reviewers for a specific post. As we mentioned, this curation can affect the quality of the peer assessment process and how students perceive it. Therefore, we defined two main research questions:}

\vskip .3cm

\textcolor{black}{\textit{RQ1: Do self-reported social relationships affect the peer assessment quality in an e-learning course?}}

\textcolor{black}{\textit{RQ2: Do students perceive social relationships bias the peer assessment process in an e-learning course?}}

\vskip .5cm

%Therefore, we define our research question as: \emph{Do self-reported social relationships affect the peer assessment process in an e-learning course?}

\textcolor{black}{We analyzed the quality of the peer assessment environment through its accuracy and fairness. We also employed questionnaires to collect user perceptions. All collected data originates from a university course that supports e-learning and peer assessment, as presented in the next section.}
%\todo[inline]{RV1: Add the research design in this section. Here, one research question is stated, but based on the results, which include peer assessment findings and students' perceptions, you have to add another question.}

\subsection{E-learning course}
%How PCM works.
We leverage a course named Multimedia Content Production (MCP), from the MSc in Information Systems and Computer Engineering at Instituto Superior Técnico — Universidade de Lisboa, which uses Moodle\footnote{Moodle is a virtual learning environment that enables you to create powerful, flexible, and engaging online learning experiences \citep{rice_moodle_2006}. At \url{www.moodle.org}.} as its virtual learning environment. The Moodle platform enables students to obtain all the necessary content to work successfully in lectures, discuss any topics related to the course curriculum, and submit assignments to be assessed by the professors. 
\textcolor{black}{There is already an extensive body of research based on this course (e.g., \citep{Alves2023}, \citep{nabizadeh2021}, \citep{barata2017} and \citep{Barata2014}).}

\begin{figure}[htb]
\centering
\includegraphics[width=0.7\textwidth]{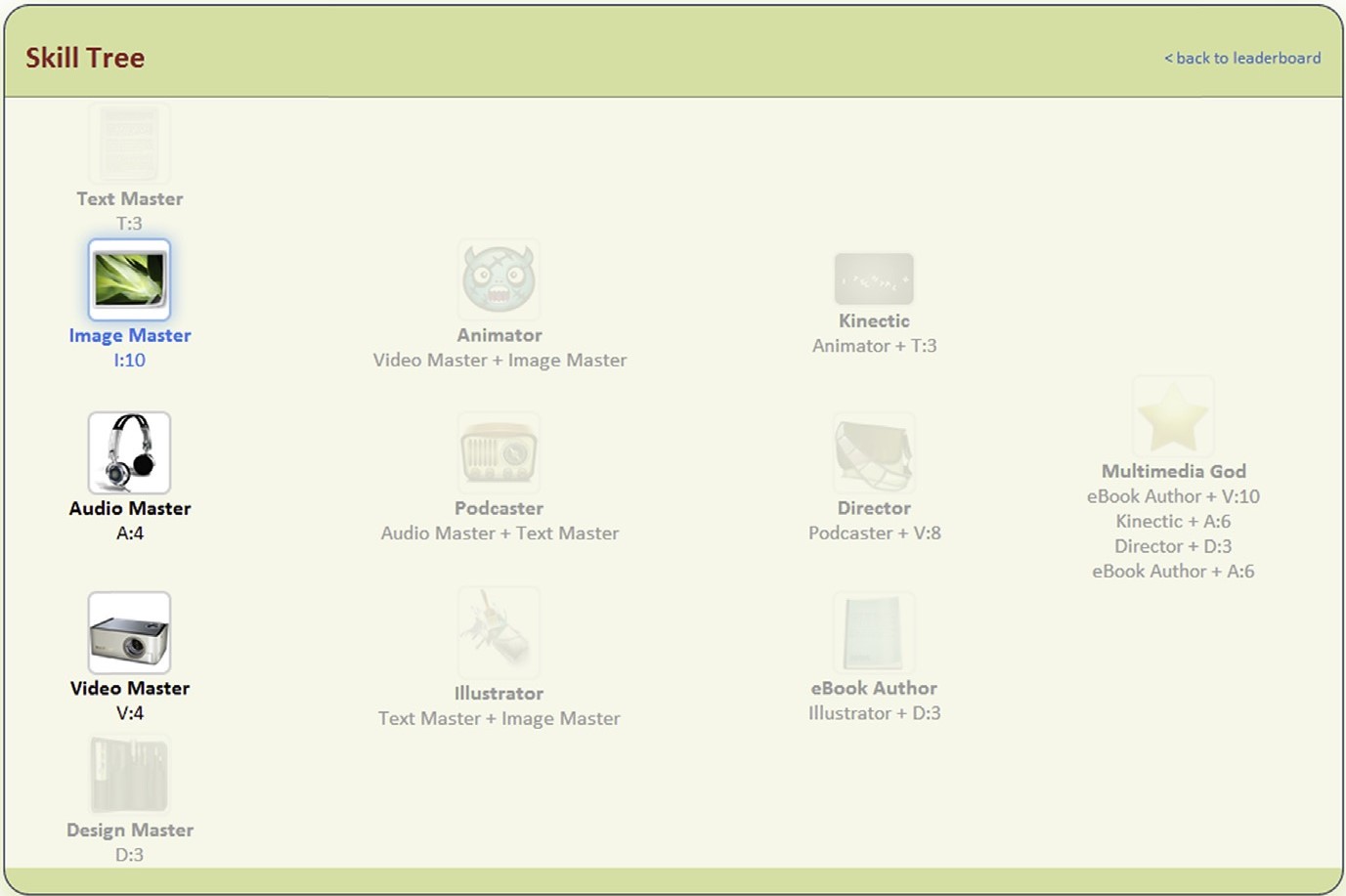}
\caption{Visual depiction of the Skill Tree used in MCP.}
\label{fig:skilltree}
\end{figure}

Students typically have two weekly lectures and one laboratory class. Although MCP is traditionally a blended learning course, the faculty ran it exceptionally in a remote setting due to the COVID-19 pandemic. In this case, students attended theoretical lectures and practical laboratories through a video-conference platform. The \textit{skill tree} is among the different grading components.
This element focuses on producing several types of multimedia content throughout the semester (Figure \ref{fig:skilltree}).
In addition to class exercises, students must deliver and have more than average grades in the \textit{skill tree}.
This strategy allows students to obtain a final high mark.
Students submit \textit{skill tree} exercises (\textit{skills}) to Moodle Forums for reviewing and assessment by professors.
There, professors review and assess the skills on a 6-point scale and some written feedback.
Afterwards, the student can reply with an improved version based on the critique if the minimal acceptance requirements were unmet or if they wish to improve their exercise's quality (and rating).
We took advantage of these Moodle Forums and submissions to have peers assess and give feedback to each other's \textit{skill tree} submissions using a Moodle plugin.

\subsection{Peer Assessment Environment}
%How the plugin works: forum, assign with bias, post, nominations, rating, training, notifications.
%\todo[inline, color=green]{DONE RV2(4): In 3.2 you mention that posts were assessed according to the recommendations found in the literature. What were these recommendations exact?}

%\todo[inline]{RV2(5): I previously asked a question about the assessment rubric for written feedback. What does this rubric look like?}

%\todo[inline]{RV2(6): Related to this, you mention that after three assessments had been done, the information would also become available to students. However, how can you guarantee that the 4th or 5th assessment was independent? Especially since students may also have talked to each other about this?}

We created PeerForum, a Moodle plugin to enable peer assessment of students'  \textit{skill tree} submissions. As part of the functionality provided by the plugin, at the beginning of the semester, students had to answer a \textcolor{black}{two-item survey} where they nominated the peers they liked the most and the peers who they liked the least \citep{coie_dimensions_1982}. At least four peers per category (liked and disliked) were required to be nominated by each student. After selecting peers, all enrolled students could reply to the professor's post with their solution for the \textit{skill}. \textcolor{black}{The plugin assigned each post to five students for assessment. We chose this amount according to \cite{luo_peer_2014} and \cite{cho_validity_2006}, who recommended using between three and five graders and four to six graders, respectively.} This assignment was not wholly random, consisting of a weighted sum of peers who nominated the post's author (if there were enough) and peers who did not. This strategy allowed for more data to be analyzed.
Each of these five students received a Moodle notification and had 48 hours to complete the peer assessment until it expired.

\begin{figure}[htb]
\centering
\includegraphics[width=\textwidth]{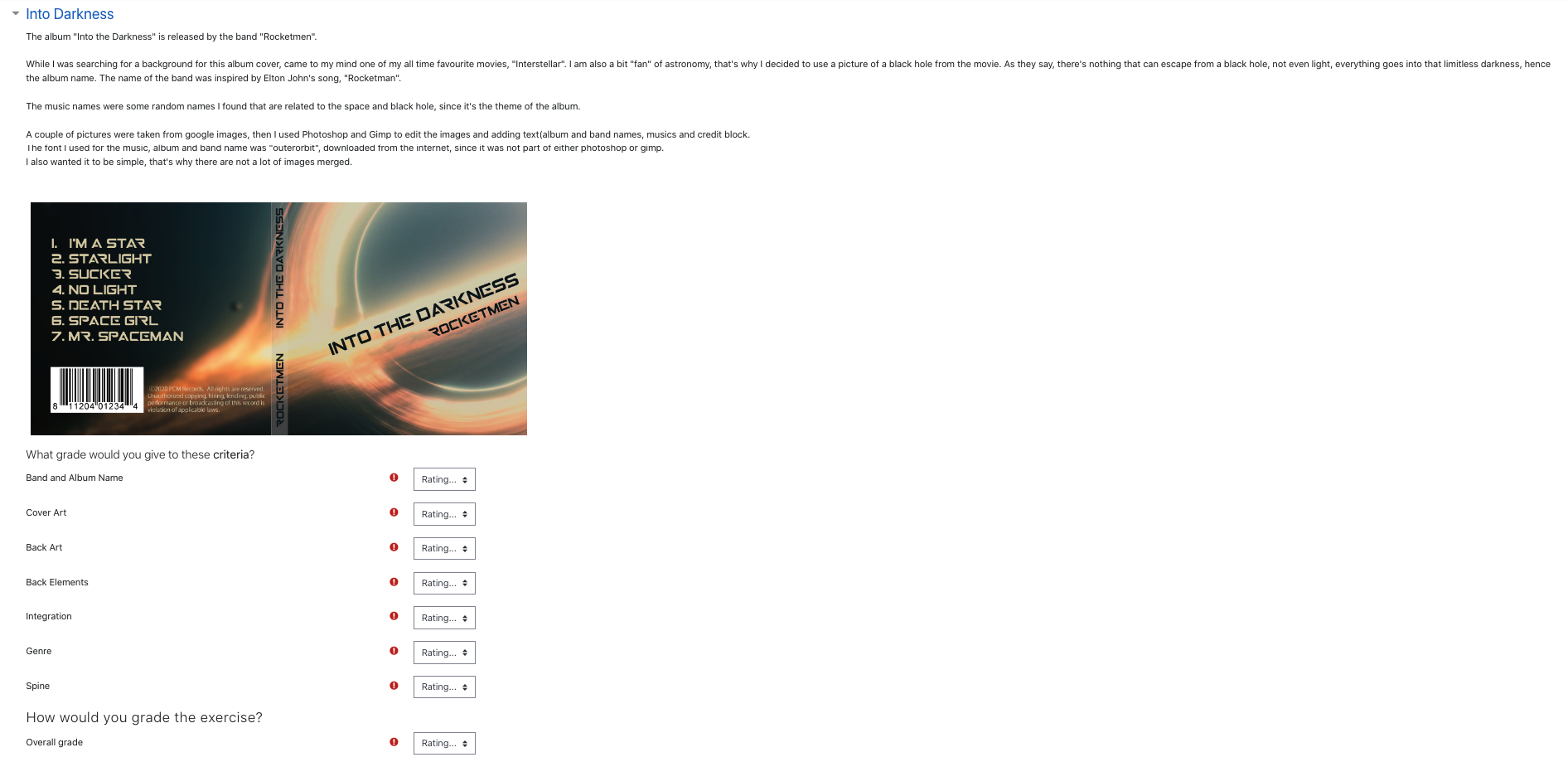}
\caption{PeerForum screen of an example of a training post.}
\label{fig:training_post}
\end{figure}

Professors also created a training page for each \textit{skill} for students to practice the peer assessment and get familiar with examples and exercises.
First, each training page contained an example of the multimedia piece expected to complete the skill and which criteria the faculty used to grade the students' submissions.
Then, the page contained at least two sample submissions for the students to grade: skill tree submissions from previous years \textcolor{black}{and curated by the faculty.
The page required students to grade between 0 (\textit{the post does not meet the requirements}) and 5 (\textit{excellent}), each sample following the criteria previously presented (see Figure~\ref{fig:training_post}).
We did not train or process the written feedback that the students provided to the post.
However, the faculty trained students to fill out these rubrics in the theoretical classes.
We considered that the student completed their training only when the grades the student provided matched exactly the grades the faculty provided for each sample.}
Furthermore, students could only submit their work for peer assessment after completing the training page.
This approach enforced that students had the same baseline on how to grade and acknowledged which criteria were more relevant for each skill, thus preparing them for upcoming peer assessment assignments and helping them produce submissions to the skill better aligned with the course's objectives.

\begin{figure}[htb]
\centering
\includegraphics[width=0.7\textwidth]{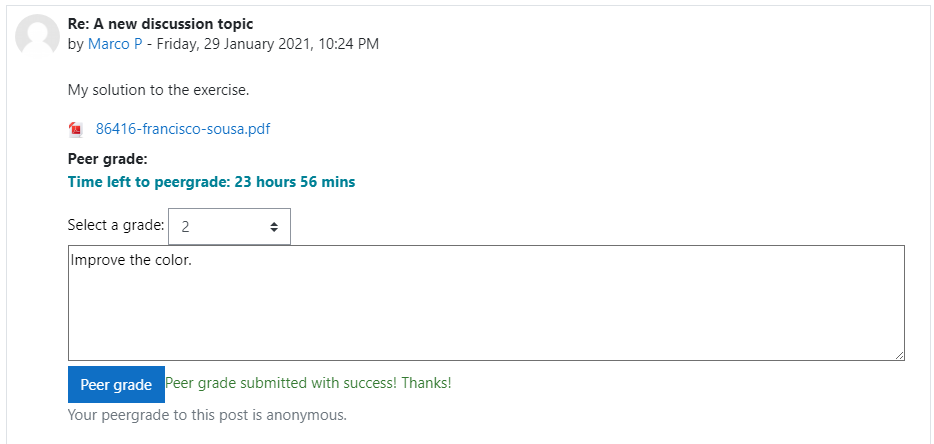}
\caption{PeerForum screen of a submission being assessed.}
\label{fig:pfpostinpgstdassignedfill}
\end{figure}

Upon receiving a peer assessment assignment, we required students to complete the training exercises from the corresponding \textit{skill}, and only then would they be allowed to perform the peer assessment: evaluate the exercise by submitting a grade (peer grade) and some written feedback (Figure \ref{fig:pfpostinpgstdassignedfill}). Besides the five student graders, the professor also graded (rating) and gave feedback on each post by making a post reply. Both grades were on a 6-point scale (0 to 5), with clearly defined criteria for each point. \textcolor{black}{A well-defined and extensive scale allows us to more effectively identify subtle differences} in grading (\citep{barata2017}).

During the peer assessment process of each post, all the submitted peer grades, ratings, feedback, and professor replies were hidden from students until the post reached a minimum number of three peer assessments, a number found reasonable in the related work. At that point, all students could see the professor's rating and reply for each post and the average of all peer grades (final peer grade), except those who could still assess the post or until the 48 hours expired. Individual peer grades and feedback were only visible to the post's author, but each grader's identity remained anonymous.
The student author could improve their work and resubmit it for appraisal based on the feedback. The original peer graders, already familiar with the work and able to better understand those improvements, would, in that case, be assigned to the resubmission post.
If the algorithm selected a student to assess a post of a colleague not present in their nominations list, we later asked them to rate that colleague on a 6-point scale (0 to 5, where 0 signifies not being acquainted with that colleague).

\subsection{Data Collection and Processing}\label{data-col}
%How the plugin was used in PCM, when n and with how many students, what data was collected, and what it means.

%\todo[inline]{RV2(7): How many of the final set of 64 respondents had also filled out the final questionnaire?}

%\todo[inline]{RV2(8): Did you also check whether professors approached the assessment similarly?}

The PeerForum plugin replaced the current Moodle Forum in the MCP course 2021. In this edition, 69 students completed the course. By the end of the semester, we also presented them with a final questionnaire we developed on their opinion on peer assessment. The students were asked to rate, on a 5-point Likert scale, their thoughts on peer grading usefulness, fairness, and difficulty before and after the course; the quality of the feedback provided and the attention given to it; the perceived amount of effect the students' relationships had on the peer grades given and received; the usefulness of the training pages in peer grading and completing the skills; and the usefulness of peer grading in achieving the skills. A text box was also open to general feedback and suggestions for course improvement.
Although the literature has shown that peer assessment works best when mandatory \citep{he_peer_2015}, we could not implement that condition in MCP due to school policies, so we awarded extra grade points for participation.
\textcolor{black}{The scale had a high level of internal consistency, as determined by a Cronbach's alpha of 0.958.}

There were 26 \textit{skills} with submissions, counting 863 posts with at least one peer assessment and 625 posts peer-assessed at least three times. The 77 students who nominated peers summed up to a total of 699 nominations and performed 1144 peer ratings. Only 56 students answered the final questionnaire.
Regarding faculty members, eight professors oversaw grading the skill tree. Each was solely responsible for a subset of skills, thus ensuring consistency of criteria in evaluating skills.

We removed the students \textcolor{black}{who never assessed} and ended up with a final set of 64 students \textcolor{black}{(45 men, 17 women, and two identifying with another gender) aged 22 ± 1.61 years old}. Moving forward, we will only analyze peer assessments from these students. Then, we removed from the dataset all peer assessments that contained no feedback related to the post, leaving 2681. The final post's peer grade is the rounded average of all peer grades for each post. For each peer assessment, the relationship between the post's author and the peer grader depends on whether the first was peer-nominated or peer-rated by the second.

\section{Data Analysis and Results}
This section describes the results of the experiment. It starts by analyzing the peer assessment fairness and accuracy and then the impact of the students' relationships on the metrics. We also discuss our results and state the limitations of our study.

\subsection{Fairness}
%ICC;
Fairness measures the consistency of scores given by multiple student graders, reflecting the general agreement among the students assigned to assess the same post -- inter-rater reliability.
Since the algorithm selected peer graders from a larger pool of enrolled students and a different set of students assessed each post, we measured their agreement with form 1 of the Intraclass Correlation Coefficient [ICC(1)].

The guidelines suggested in \cite{koo_guideline_2016} will be used to interpret the ICC results: values lower than .50 are indicative of poor reliability, values between .50 and .75 indicate moderate reliability, between .75 and .90 indicate good reliability and values greater than .90 mean excellent reliability.

%ICC for several n grades;
The ICC(1) test was applied to the data obtained throughout our study. This test can only be made to data sets with the same number of item ratings. Hence, we grouped posts by their number of completed peer assessments. ICC estimates and their 95\% confidence intervals (CI) were calculated using SPSS statistical package version 26 (SPSS Inc, Chicago, IL) based on single and average ratings, absolute agreement, one-way random-effects model (see Table \ref{tab:fairiccbynofpgs}).

\begin{table}[htb]
\caption{ICC estimates and their 95\% CI by number of peer assessments.}
\centering
\begin{tabular}{ccrrr}
\multirow{2}{*}{\textbf{\begin{tabular}[c]{@{}c@{}}No. of Peer\\ Assessments\end{tabular}}} & \multirow{2}{*}{} & \multirow{2}{*}{\textbf{\begin{tabular}[c]{@{}r@{}}Intraclass\\ Correlation\end{tabular}}} & \multicolumn{2}{c}{\textbf{95\% CI}} \\
 &  &  & Lower Bound & Upper Bound \\ \hline
\multirow{2}{*}{\begin{tabular}[c]{@{}c@{}}2\\ (N = 124)\end{tabular}} & \multicolumn{1}{c|}{\textbf{Single Measures}} & .188 & .013 & .352 \\
 & \multicolumn{1}{c|}{\textbf{Average Measures}} & .316 & .026 & .520 \\ \hline
\multirow{2}{*}{\begin{tabular}[c]{@{}c@{}}3\\ (N = 199)\end{tabular}} & \multicolumn{1}{c|}{\textbf{Single Measures}} & .435 & .350 & .519 \\
 & \multicolumn{1}{c|}{\textbf{Average Measures}} & .698 & .617 & .764 \\ \hline
\multirow{2}{*}{\begin{tabular}[c]{@{}c@{}}4\\ (N = 203)\end{tabular}} & \multicolumn{1}{c|}{\textbf{Single Measures}} & .238 & .168 & .315 \\
 & \multicolumn{1}{c|}{\textbf{Average Measures}} & .556 & .447 & .648 \\ \hline
\multirow{2}{*}{\begin{tabular}[c]{@{}c@{}}5\\ (N = 186)\end{tabular}} & \multicolumn{1}{c|}{\textbf{Single Measures}} & .222 & .159 & .293 \\
 & \multicolumn{1}{c|}{\textbf{Average Measures}} & .588 & .486 & .675
\end{tabular}
\label{tab:fairiccbynofpgs}
\end{table}

The ICC Single Measures estimates the reliability of each of the randomly selected student graders when grading the same assignment. For posts with five peer assessments, the coefficient value of .222, with a 95\% CI of \([.159, .293]\), is considered poor in strength by the guidelines above, suggesting peer grades vary significantly among individual student graders and a single student's grade is not very reliable.
The ICC Average Measures of .588, with 95\% CI in \([.486, .675]\), shows poor to mostly moderate reliability.
In our use case, we plan to use the average peer grade of the five students as the assessment basis instead of one. Hence, we should use the ICC Average Measures as the measurement index.
According to the results, there is barely any difference between having four and five peer assessments for one post. However, both ICC Single and Average Measures stand out when considering posts with only three peer assessments, reflecting consistently moderate reliability, almost good, overall higher than the four and five values. On the contrary, the ICC results drop to their lowest value when checking posts with only two peer assessments.

This result confirms what has also been observed in [Blind for Review]: three peer assessments are usually enough for students to reach an agreement on the post's final peer grade. This is the minimum number of assessments needed to consider the peer assessment activity sufficient and display the results of both it and the professor rating.

\subsection{Accuracy}
%Pearson with final pg;
In this study, we used the professor's grade as the ground truth on whether the final peer grade of a post is accurate: the professor rating. The means and standard deviations of the posts' metrics are in Table \ref{tab:accumeanmetrics}. Accuracy measures the similarity between the final peer grade and the professor's grade in the same post by assuming professors award fair and accurate scores -- a convergent validity.

\begin{table}[htb]
\caption{Mean and standard deviations of post's metrics: rating, average of all peer grades and its rounded value (final peer grade). N = 588.}
\centering
\begin{tabular}{crrr}
\multicolumn{1}{l}{} & \textbf{\begin{tabular}[c]{@{}r@{}}Professor\\ Rating\end{tabular}} & \textbf{\begin{tabular}[c]{@{}r@{}}Final Peer\\ Grade\end{tabular}} & \textbf{\begin{tabular}[c]{@{}r@{}}Average of\\ Peer Grades\end{tabular}} \\ \cline{2-4} 
\multicolumn{1}{c|}{\textbf{Mean}} & 3.16 & 3.62 & 3.58 \\
\multicolumn{1}{c|}{\textbf{\begin{tabular}[c]{@{}c@{}}Standard\\ Deviation\end{tabular}}} & .92 & .61 & .57
\end{tabular}
\label{tab:accumeanmetrics}
\end{table}

\begin{figure}[htbp]
	\centering
	\subfigure[final peer grade.]{\label{fig:accurfpgloess} 		\includegraphics[width=0.45\textwidth]{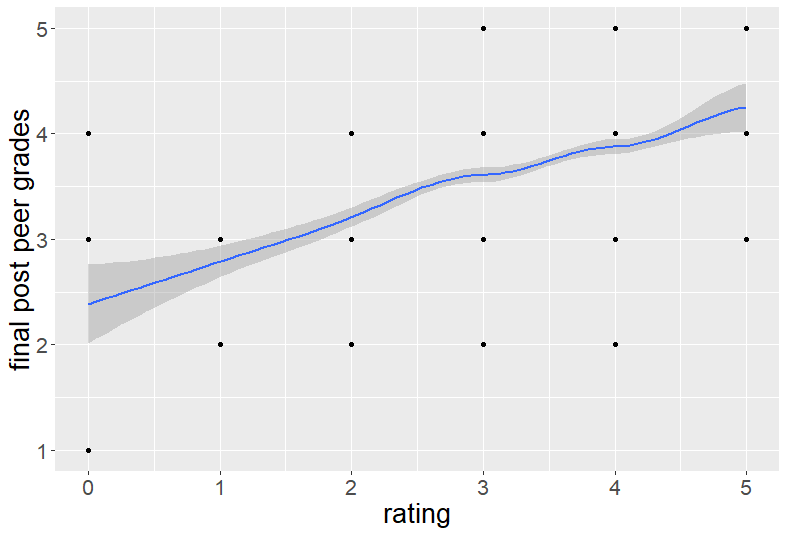}} \qquad
	\subfigure[average of peer grades.]{\label{fig:accuravgpgloess}
		\includegraphics[width=0.45\textwidth]{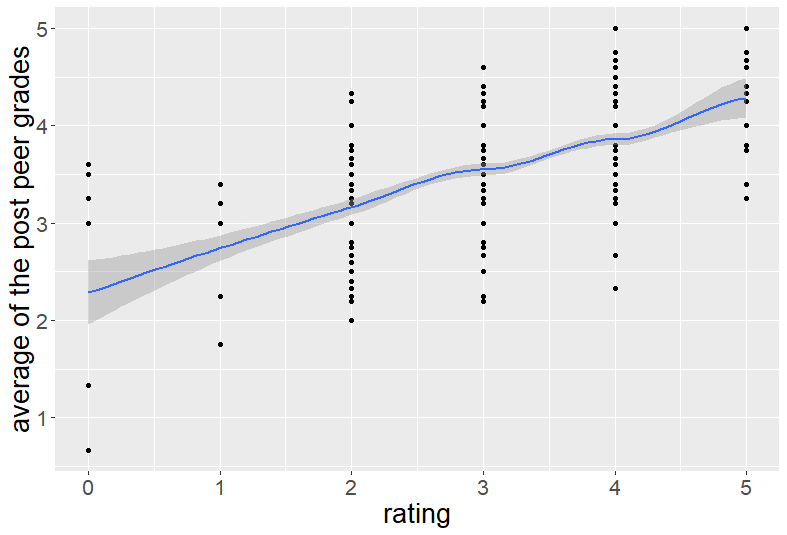}}
	\caption{Correlation between the posts' ratings and the peer grades.}
	\label{fig:accurpgloess}
\end{figure}

Since submission grades are ordinal, we assessed the relationship between each post's professor grade and final peer grade with a Spearman's rank-order correlation.
Five hundred eighty-eight posts were considered, each with at least three peer assessments.
Preliminary analysis showed the relationship to be monotonic, as assessed by visual inspection of a scatterplot.
There was a statistically significant moderate positive correlation between each post's professor rating and the final peer grade, \(r_s\)(586) = .49, \(p\) $<$ .001 (Figure \ref{fig:accurfpgloess}). When considering the average of the peer grades (without rounding), the correlation becomes stronger, with \(r_s\)(586) = .56, \(p\) $<$ .001 (Figure \ref{fig:accuravgpgloess}).

%Pearson with several n grades;
The minimum number of peer assessments required to be considered for a valid final peer grade was studied by looking at its impact on accuracy. Table \ref{tab:accuspearmanrfpgradebymingrades} shows Spearman's Correlation Coefficient \(r_s\) and its significance \(p\)-value between the professor rating and average of peer grades of posts with different minimum peer assessments. The correlation accuracy pattern seems consistent with the one observed in the reliability ICC (Table \ref{tab:fairiccbynofpgs}): there is a strong and significant correlation in all cases; having at least three peer assessments is recommended to obtain the most accuracy when comparing with the professor grade; \textcolor{black}{and have five peer assessments is the ideal}. These results confirm the literature findings that peer assessment is accurate: students can overall give a grade like the professor's, even if all are slightly higher on average (Table \ref{tab:accumeanmetrics}).

\begin{table}[htb]
\caption{Spearman's Correlation Coefficient and significance value between the professor rating and the average post peer grades by a minimum number of peer assessments.}
\centering
\begin{tabular}{lrrrrr}
 & \multicolumn{5}{c}{\textbf{\begin{tabular}[c]{@{}c@{}}Minimum number of\\ Peer Assessments in post\end{tabular}}} \\
\textbf{} & \begin{tabular}[c]{@{}r@{}}5\\ (N = 186)\end{tabular} & \begin{tabular}[c]{@{}r@{}}4\\ (N = 389)\end{tabular} & \begin{tabular}[c]{@{}r@{}}3\\ (N = 588)\end{tabular} & \begin{tabular}[c]{@{}r@{}}2\\ (N = 712)\end{tabular} & \begin{tabular}[c]{@{}r@{}}1\\ (N = 806)\end{tabular} \\ \cline{2-6} 
\multicolumn{1}{l|}{\textbf{\(r_s\)}} & .596 & .549 & .558 & .545 & .512 \\
\multicolumn{1}{l|}{\textbf{\(p\)}} & \textless .001 & \textless .001 & \textless .001 & \textless .001 & \textless .001
\end{tabular}
\label{tab:accuspearmanrfpgradebymingrades}
\end{table}

We also compared the difference between professor ratings and students' peer grades grouped by the post's rating to check which latter was more (and less) in line with the student's grades. The results in Table \ref{tab:accusameasrating} show that students tend to be more positive and moderate than professors when giving their assessments: in posts with negative grades (below 3), students award more points, rarely giving the complete fail. In posts with 3, students give them 4. Students only agree (strongly) with professors when giving 4's. As for the maximum grade, students usually choose it fewer times than expected.

\begin{table}[htb]
\caption{Final peer grade difference from rating depending on the post rating}
\centering
\resizebox{\textwidth}{!}{%
\begin{tabular}{clrrrrrrrrr}
\multicolumn{1}{l}{} &  & \multicolumn{9}{c}{\textbf{Final Peer Grade difference from Rating}} \\
\multicolumn{1}{r}{} & \multicolumn{1}{r}{} & -4 & -3 & -2 & -1 & 0 & 1 & 2 & 3 & 4 \\ \cline{3-11} 
\multirow{7}{*}{\textbf{\begin{tabular}[c]{@{}c@{}}Post\\ Rating\end{tabular}}} & \multicolumn{1}{l|}{0} & - & - & - & - & 0 (0,0\%) & \textbf{3 (42,9\%)} & 0 (0,0\%) & 2 (28,6\%) & 2 (28,6\%) \\
 & \multicolumn{1}{l|}{1} & - & - & - & 0 (0,0\%) & 0 (0,0\%) & 2 (40,0\%) & \textbf{3 (60,0\%)} & 0 (0,0\%) & 0 (0,0\%) \\
 & \multicolumn{1}{l|}{2} & - & - & 0 (0,0\%) & 0 (0,0\%) & 10 (7,8\%) & \textbf{82 (63,6\%)} & 37 (28,7\%) & 0 (0,0\%) & - \\
 & \multicolumn{1}{l|}{3} & - & 0 (0,0\%) & 0 (0,0\%) & 2 (0,9\%) & 80 (37,6\%) & \textbf{130 (61,0\%)} & 1 (0,5\%) & - & - \\
 & \multicolumn{1}{l|}{4} & 0 (0,0\%) & 0 (0,0\%) & 1 (0,5\%) & 31 (14,5\%) & \textbf{175 (81,8\%)} & 7 (3,3\%) & - & - & - \\
 & \multicolumn{1}{l|}{5} & 0 (0,0\%) & 0 (0,0\%) & 1 (10,0\%) & \textbf{11 (55,0\%)} & 7 (35,0\%) & - & - & - & - \\
 & \multicolumn{1}{l|}{\textbf{Total}} & 0 (0,0\%) & 0 (0,0\%) & 3 (0,5\%) & 44 (7,5\%) & \textbf{272 (46,3\%)} & 224 (38,1\%) & 41 (7,0\%) & 2 (0,3\%) & 2 (0,3\%)
\end{tabular}%
}
\label{tab:accusameasrating}
\end{table}

\subsection{Relationships}
%ANOVA

The main focus of this study is to infer whether the personal relationships between students impact their assessments. Hence, we collected some existing relationships (like or dislike) and compared them to a control group of neutral acquaintances. We did not analyze the peer assessments in which the relationship between the post's author and the student grader is unknown.

From 2681 peer assessments, we know the relationships between post author and peer grader of 2561 (95.5\%), but only 2238 (83.5\%) are in posts with at least three peer assessments.
Of these, 684 (30.6\%) are between students who like each other, 152 (6.8\%) are between students who do not like each other, and the rest, 1402 (62.6\%) report they have neutral feelings towards the author of the post.

We conducted two one-way Welch ANOVAs to study the effect of student relationships on the fairness and accuracy of peer assessment, determining if the peer grade difference from the final post grade and professor rating was distinct for different relationship types (like, dislike, neutral).

\begin{figure}[htbp]
	\centering
	\subfigure[final peer grade]{\label{fig:relasdifffpg} 		\includegraphics[width=0.45\textwidth]{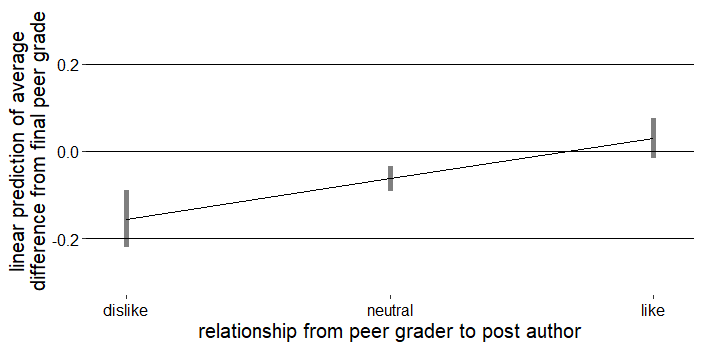}} \qquad
	\subfigure[professor rating]{\label{fig:relasdiffrating}
		\includegraphics[width=0.45\textwidth]{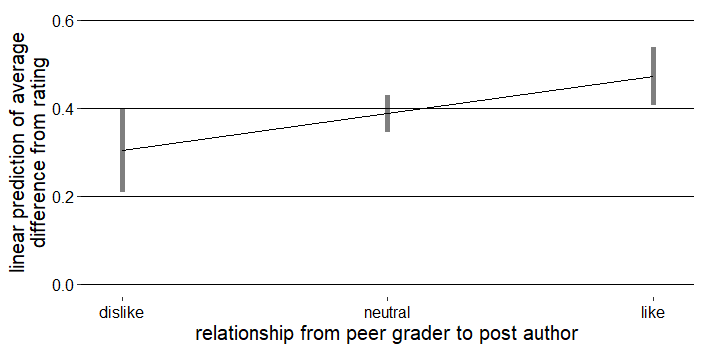}}
	\caption{Linear prediction of the differences from grades by the relationship between students.}
	\label{fig:relasdiff}
\end{figure}

Both the final peer grade and rating differences were statistically significantly different between the three relationship types, Welch's \(F\)(2, 412.373) = 7.472, \(p\) = .001, Welch's \(F\)(2, 411.337) = 3.561, \(p\) = .029, respectively.
The peer grade difference from the final post's peer grade increased from the dislike group (M = -.12, SD = .597) to the neutral (M = -.07, SD = .639) and then to the like group (M = .04, SD = .69). The same for the peer grade difference from professor rating: increased from the dislike group (M = .36, SD = .881) to the neutral (M = .37, SD = .931) and then to the like group (M = .49, SD = .987).
Games-Howell post hoc analysis revealed that the mean increase from neutral to like was statistically significant in both the final peer grade difference (.111, 95\% CI \([.04, .19]\), \(p\) = .001) and the rating difference (.118, 95\% CI \([.01, .22]\), \(p\) = .025), as well as the increase from increase from dislike to like (.156, 95\% CI \([.03, .20]\), \(p\) = .014) in final peer grade difference.

Although students who like their peers tend to give them higher grades, the mean difference is meager, reaching much less than half a point (Figure \ref{fig:relasdiff}). This analysis shows that although there is a trend in students grading peers they like with higher grades, the difference and probability are not strong enough to impact the post's final grade for our scale. The same conclusions would not be applicable in a high-stakes assessment with a larger scale, as this discrepancy could be more accentuated and meaningful for the final grade.

\subsection{Student Perceptions}
At the end of the semester, we asked students to reply to a questionnaire about the course, with some questions regarding peer assessment. \textcolor{black}{Of the 56 students who responded, only 53 had peer-graded any post and were part of the %set of 
64 students mentioned in \ref{data-col}. The following analysis only takes into consideration the replies from those 53 students. From that final set of students, 27 (50.9\%) had already heard of peer assessment before the course, and 21 (39.6\%) had already peer-graded someone.
When asked if they thought peer assessment was helpful before the course, 23 agreed (43.4\%), and 13 (24.5\%) disagreed; after, 29 (54.7\%) agreed, with 10 changing their opinion for the better.
When asked about fairness, 13 students changed their opinion for the better compared to what they thought before the course, now with 28 (52.8\%) agreeing and 11 (20.8\%) disagreeing that peer assessment is fair.
At the end of the course, 16 students (30.2\%) agreed, and 25 (47.2\%) disagreed that peer assessment is difficult, as 11 lowered their agreement after.}

\textcolor{black}{Regarding the impact and quality of the given feedback, although a great portion of students (45.3\%) agree/completely agree their feedback was heard, many (37.7\%) are indecisive. Similarly, many students agree (47.2\%) that they provided quality feedback, although only 15.1\% completely agree with the statement (26.1\% are indecisive). Finally, around 11.3\% completely disagree, and 34\% disagree with the statement, ''I think peer assessing my colleague's submissions did not help me when completing the \textit{skills}.'' At the same time, 17\% of students completely agree with it.}

\textcolor{black}{Regarding their perceptions of relationships, students disagree (75.5\%) that their relationship with their colleagues affected the assessment they gave them. Only six students (11.3\%) agreed. The reverse is similar: many students (67.9\%) disagree that their relations with them influenced the feedback they received.
Students have confidence in their peers' grades, with 49.1\% disagreeing/completely disagreeing that their colleagues are bad at evaluating compared to the professors. 34\% are neutral.}

\textcolor{black}{Furthermore, in the open text boxes, most mentions to peer grading praised the feature.
From the 53 replies, eight students mentioned peer grading and the peer grading extra grade points as the most effective/interesting achievement from the course. It was because they could ``give feedback to their colleagues' work,'' ``learn about the skills (some before submitting work to them) and what the teachers expected.'' It helped students improve their work and made them more engaged in the course by making them ``keep up with Moodle'' and be attentive to the skills and how their peers were doing in each.}

\textcolor{black}{They were motivated to collaborate, engage in the course, and ``give good and fast feedback to help colleagues.'' Students also reported that peer grading was a ``useful feature'' as it helped them learn and fix things the professor missed, but their colleagues pointed that out. Two other students suggested that the peer grade should have a bigger impact on the student's final grade, meaning they find peer grading fair.}

\textcolor{black}{Additionally, five students pointed out peer grading should be mandatory because it is an ''important process'' and it ''helps students improve their work'', both by ''the possibly good feedback I might give my peers'', as well as because ''they would get to see the professors' feedback sooner''.}

\textcolor{black}{Only one student suggested that the person being graded should always be able to see the professor's reply. The remaining students did not report issues with having to wait for the peer grades before having access to the professor's reply/grade, although one pointed out this was a problem at the beginning of the semester and it improved throughout; others noted that this also worsened in the final days of the semester when students were focusing on delivering last minute work.}

\textcolor{black}{Three other students mentioned the quality of the written feedback in some peer grades, suggesting that the faculty should reward thoughtful/helpful feedback. They even proposed a mechanism to rate the feedback.
No student said peer grading increased their workload, being classified as ``easy enough work for a good chunk of credits,'' taking ``not even 5 minutes of the day,'' and ``offering some utility to the rest of the students.''}

\textcolor{black}{Only one student who peer graded said it was their least effective/interesting achievement from the course. They said they were ``unsure what to write because the examples given in some of the training pages were insufficient to completely understand the criteria for a skill.'' The same student added that their feedback consisted of small ``nice work'' variations to unblock their peers' grades and gain bonus points.}

\textcolor{black}{Finally, from the three students who replied to the final questionnaire and whose replies we did not consider in the above analysis, two mentioned peer grading as their most ineffective/uninteresting course achievement because, in their opinion, ``it helps nothing to the experience and just slows down the turnaround time to receive a teacher's grade.'' Their replies were not included in the remaining data because they did not grade any posts.}

\section{Discussion}
%Peer assessment is fair and accurate like literature says and students like it;

%\todo[inline]{RV2(9): The starting point in this study is peer relationships, but how well do students actually know each other in an online environment? Would you expect this to might explain why you have not found many differences?}

%\todo[inline]{RV2(12): I asked a question about motivation. What was said in the text boxes on motivation, how did you analyze and code this question? What did students say about it?}

This study aimed to understand how pre-existing personal relationships between students affect peer assessment in an e-learning environment and whether it can accurately and fairly replace traditional evaluation. 
\textcolor{black}{Regarding the quality of the peer assessment, we observed that students converge with similar grades to a specific post. In particular, our algorithm to pick the reviewers of a post according to their relationship with the post's author requires a minimum of three assessments to agree on the post's final peer grade. %This 
Our result aligns with the work of \cite{luo_peer_2014} regarding the minimal number of reviewers for rater convergence. The exception is when students need to rate posts in the extremes of the rating scale, i.e., negative grades and excellent grades. For instance, we found that these posts generated an average standard deviation of all the peer grades larger than one point in posts with a final peer grade of one. %This result 
Our finding suggests that the rubrics were not precise enough to capture the worst and the best posts' grading features. Similar to \citet{panadero2013impact}, our results showcase the limitations of not using fine-tuned rubrics.}

\textcolor{black}{Next, we consider the professor's rating as the reference point for each post's grade. Although students tend to grade slightly higher than the professors, it is at most half a point on average. Considering that reviewers grade with discrete values with a granularity of one, we assume that students give peer ratings similar to those of an instructor. In particular, we found a strong correlation between the professor's grade and the final peer grade of each post with at least three peer assessments. Contrary to recent meta-analyses (e.g., \citet{panadero_empirical_2019,li2016peer}), our results are in line with previous literature suggesting that anonymous groups tend to provide more accurate grades (e.g.,~\citet{azarnoosh_peer_2013,luo_peer_2014,he_peer_2015, guler2017use,usher_peer_2018}).
%Students are more moderate than professors when grading, like with fairness.
Moreover, only a few students had been peer-assessed before taking the course. Students reported that the peer assessment process was fair and valuable. They were also confident in their peers' grades compared to the professors' and stated it has helped them complete the \textit{skills}.
Similar to \citet{lin2018anonymous}, the students received the single-anonymized approach positively. However, our sample believes it was fair compared to \citet{lin2018anonymous}. We hypothesize that students believe our approach was fair since they found the peer assessment valuable and the final grades of the posts aligned with the professor's grades. \citet{kaufman2011students} reported a similar finding. In \citet{lin2018anonymous}, the usefulness of the peer feedback students received did not meet their expectations, hence the lower fairness perception.}

%Relationships have little influence when reported and students agree.
\textcolor{black}{With the overall fairness and accuracy of peer assessment confirmed, we continued by observing the effect of students' relationships on the process. Our findings show that the peer grader liking or disliking the author of the post they are assessing impacts accuracy and fairness. On average, students who like each other tend to give higher grades than the rest of the peer graders or the professor, and students who dislike each other tend to give lower grades than the rest of the peer graders or closer to the professor (which already grades lower than the final peer grade). These findings are in line with past results on the over- or under-scoring phenomena triggered by a relationship bias (e.g., \citet{harris2013opportunities,dominguez2016comparative,kilickaya2017peer,ersoz2018facebook}). However, when we consider the reviewers tailoring we applied when picking the set of reviewers for a post based on the relationships, we found that these differences can be ignored, not exceeding half a grade point. We hypothesize that the assignment algorithm, which ensures a post is only assigned to at most one student who likes the author and one who does not, helped balance the fairness of the final peer grade. Without this algorithm, the chances of a post being assigned mostly to students who know the author are low. By ensuring that the post's reviewers cover a heterogeneous range of relationship statuses regarding the reviewer, we could balance out the peer ratings and still converge them to final ratings similar to the professor's. Furthermore, students disagreed that their relationship influenced the feedback both given and received. Our results provide evidence to understand the role of relationship bias in peer assessment and how to control it.}

\subsection{Design Implications}

Based on our results, we %devised 
propose a set of implications that can be useful %for 
in designing the pipeline of peer assessment systems.

\textbf{\textit{There is no need for peer assessment to be anonymous.}} 
In \cite{usher_peer_2018}, the peer assessment was anonymous, and on-campus students still refrained from writing negative comments, contrasting with the unkind MOOC's feedback, showing students' awareness of grading peers with whom they have personal acquaintance. The authors conclude that although \cite{li2020does} observed better results in anonymous peer assessment, a bias is present even without knowledge of the identity of the exercise's author. \textcolor{black}{Therefore, solely relying on anonymity may not improve the peer assessment process. Moreover, recent meta-analyses (e.g., \citet{panadero_empirical_2019}) suggest that non-anonymity is better for increasing students' peer grading accuracy when compared to teachers' assessment. Our study consisted of a mix of interactions between students who knew and did not know each other, and we focused on studying the bias of these relationships. Our results do not support the idea that anonymity is also necessary if we consider the students' relationships and curate the set of reviewers for a post based on the author. The overall peer assessment was accurate, and when analyzing the grades of students who knew each other, the difference was not significantly different from those who did not. Based on these results, we do not recommend anonymizing the peer assessment.}

%Nevertheless, even without anonymity, the overall peer assessment was accurate, and when analyzing the grades between students who knew each other, the difference was not significant with these who did not. Based on these results, we do not recommend anonymizing the peer assessment.

\textbf{\textit{There should be prior training for peer assessment.}} Students are not experienced graders, nor are they fluent in the subjects being evaluated when assessing peers, which influences the accuracy and fairness of peer assessment. For example, in \cite{luo_peer_2014}, the grading criterion with the lowest peer agreement is a criterion more related to the course content, being more affected by different students' levels of prior knowledge and learning outcomes; the prior training of the grader is the most decisive factor that explains the variation of the peer assessment in \cite{li2020does}; and \cite{azarnoosh_peer_2013} specifically accredit the high agreement between professor and peer assessments to the training and practice sessions before the actual peer assessment experience, as well as the usage of clear scoring criteria. \textcolor{black}{In our study, we observed that the rubrics were useful to train students to grade a post except if it had poor or excellent quality. \citet{panadero2013impact} also leveraged rubrics to counter the relationship bias. All findings point towards the importance of creating fine-tuned rubrics and training students with them to improve the results of peer assessment.}

\textbf{\textit{\textcolor{black}{The relationship bias can be controlled}.}} Our results show that the fairness and accuracy of peer assessment are the highest for posts with at least three peer assessments, confirming the literature findings that three should be the minimum of peer assessments required for a fair and accurate final peer grade \citep{luo_peer_2014}.
\textcolor{black}{To reach this minimum of three graders, we advise future researchers to use peer nominations and ratings of student relationships to tailor the reviewers' choice for a post based on the author's relationships. Our approach shows that we could balance out the sample of reviewers and still provide grades close to the professors' ratings.}

\subsection{Limitations and Future Work}

%Future work.
\textcolor{black}{Regarding the analysis, we collected a substantial data trove from relevant student and educator interactions for future work.
The MCP course is a Master's course that usually follows a blended learning approach, with face-to-face interactions and a balanced amount of students who
know each other from previous/simultaneous courses and new students. Due to exceptional reasons, the semester in which we ran the experiment was completely online and at a distance, which may have reduced the number of relationships between students.
Additionally, several plausible data combinations were not checked and thoroughly analyzed to maintain focus on the intended objective and for brevity. Hence, more levels and aggregations of the collected data can be explored, including the evolution through the semester and questionnaire comparisons, among others.
Another limitation is using a grading scale with few levels (the 6-point scale), not allowing for a subtle distinction between grades, increasing the difference in results of the final peer grade and the professor rating.}

\textcolor{black}{Although the data set we gathered is comprehensive, there were some issues with the training variables and not an abundance of student relationships, which would benefit from a rerun of the study in a blended learning environment instead of e-learning.
For instance, the training pages were an ongoing process created throughout the semester by the faculty, with several bugs, typos, and unfinished configurations in the pages' exercises. The students eventually spotted and fixed them, but this recurrent process polluted the gathered training data, so purging was impossible.
Furthermore, the sample size in this experience does not allow us to draw substantial conclusions.
Future studies should consider increasing the number of participants to generate more data and relationship dynamics, allowing us to investigate the effect of social relationships with increased validity. Finally, our sample mainly comprises Portuguese individuals. Thus, it includes a cultural bias in our results.
}

%Have training;
%Have at least three peer grades;
%Remind students;
%Have bonus;
%No need to be anonymous.
\section{Conclusions}
%There was this problem. We did this. We concluded this. Thanks;

Peer assessment has been widely studied as a replacement for traditional evaluation, not only by reducing the professors' workload but mainly by benefiting students' \textcolor{black}{engagement} and learning. This study uses a Master's e-learning course to research the influence of pre-existing social relationships between students in peer assessment's accuracy and fairness upon prior training and through the self-collection of relationship data.
The results confirm the literature findings that peer assessment is reliable -- with students agreeing on each post's final grade -- and accurate -- students giving the same grade as the professor -- for posts with at least three peer assessments.
However, the peer grades are slightly higher.

Students' social relationships are noticeable when students who do not like the other grade their work consistently lower than students who have a positive connection. However, this difference has minimal influence on the final peer grade, and students are unaware of it. Through self-reported feedback, they agree that peer assessment is valuable and fair. It allowed them to learn throughout the semester while grading, improving their work and \textcolor{black}{engagement}.
These results allow us to conclude that peer assessment can replace traditional evaluation in an e-learning environment, as long as it is not a high-stakes assessment, benefiting students independently of their social relationships.

\bmhead{Acknowledgments}

%Omitted for the review process.
This work was supported by national funds through Fundação para a Ciência e a Tecnologia (FCT) with references SFRH/BD/144798/2019 and under project UIDB/50021/2020 (DOI:10.54499/UIDB/50021/2020).

\bibliography{sn-bibliography}% common bib file
\bibliographystyle{sn-basic}
%% If required, the content of the .bbl file can be included here once the bbl is generated
%%\input sn-article.bbl

%% Default %%
%%\input sn-sample-bib.tex%

\end{document}